%% file: main.tex
\documentclass[aps,prl,tightenlines,showpacs,floatfix,preprintnumbers,amsmath,superscriptaddress,
amssymb,amsbsy,nofootinbib,nobibnotes,twocolumn,longbibliography]{revtex4-1}

\usepackage{amsmath, amssymb, amsfonts,amsbsy,amsthm} 
\usepackage{mathtools}
\usepackage{graphicx,color} 
\usepackage{hyperref} 
\usepackage{bm}
\usepackage{times}
\usepackage{mathrsfs}
\usepackage{latexsym}
\usepackage{dsfont}
\usepackage{siunitx} 
\usepackage{hyphenat} 
\usepackage{titlesec}
\usepackage{enumitem} 
\usepackage{multirow}
\usepackage{xspace} 

\newcommand{\blue}[1]{#1}

\newcommand{\OddsSN}{\mathcal{O}^\textsc{sig}_\textsc{n}}
\newcommand{\OddsGR}{\mathcal{O}^\textsc{ngr}_\textsc{gr}}

\newcommand{\smallmathcal}[1]{{\scalebox{0.6}{$\mathcal{#1}$}}}

\input{BayesFactors.txt}
\input{peResults.txt}

\begin{document}
\title{A Search for Tensor, Vector, and Scalar Polarizations in the Stochastic Gravitational-Wave Background}
\input{LVC_Aug2017_StochasticTensorVectorScalar-prd}

\begin{abstract}
The detection of gravitational waves with Advanced LIGO and Advanced Virgo has enabled novel tests of general relativity, including direct study of the polarization of gravitational waves.
While general relativity allows for only two tensor gravitational-wave polarizations, general metric theories can additionally predict two vector and two scalar polarizations.
The polarization of gravitational waves is encoded in the spectral shape of the stochastic gravitational-wave background, formed by the superposition of cosmological and individually unresolved astrophysical sources.
Using data recorded by Advanced LIGO during its first observing run, we search for a stochastic background of generically polarized gravitational waves.
We find no evidence for a background of any polarization, and place the first direct bounds on the contributions of vector and scalar polarizations to the stochastic background.
Under log-uniform priors for the energy in each polarization, we limit the energy densities of tensor, vector, and scalar modes at 95\% credibility to $\Omega^T_0 < \num{\OmgTlogUniformTVS}$, $\Omega^V_0 < \num{\OmgVlogUniformTVS}$, and $\Omega^S_0 < \num{\OmgSlogUniformTVS}$ at a reference frequency $f_0 = 25$ Hz.
\end{abstract}

\maketitle


\textit{Introduction.} --
The direct detection of gravitational waves offers novel opportunities to test general relativity in previously unexplored regimes.
Already, the compact binary mergers \cite{GW150914,GW151226,GW170104,GW170814,GW170817} observed by Advanced LIGO (the Laser Interferometer Gravitational Wave Observatory) \cite{LIGOdetector,LIGOdetector2} and Advanced Virgo \cite{VirgoDetector} have enabled improved limits on the graviton mass, experimental measurements of post-Newtonian parameters, and inference of the speed of gravitational waves, among other tests \cite{GW150914_GR,O1BBH,GW170104,GW170817_GRB}.

Another central prediction of general relativity is the existence of only two gravitational-wave polarizations: the tensor plus and cross modes, with spatial strain tensors
	\begin{equation}
	\hat e_+ = \begin{pmatrix}
		1 & 0 & 0 \\
		0 & -1 & 0 \\
		0 & 0 & 0
		\end{pmatrix} \hspace{0.75cm}
	\hat e_\times = \begin{pmatrix}
		0 & 1 & 0 \\
		1 & 0 & 0 \\
		0 & 0 & 0
		\end{pmatrix}
	\end{equation}
(assuming waves propagating in the $+\hat z$ direction).
Generic metric theories of gravity, however, can allow for up to four additional polarizations: the $x$ and $y$ vector modes and the breathing and longitudinal scalar modes, with basis strain tensors \cite{Eardley1973a,Eardley1973b,Will2014}
	\begin{equation}
	\begin{aligned}
	\hat e_x &= \begin{pmatrix}
		0 & 0 & 1 \\
		0 & 0 & 0 \\
		1 & 0 & 0
		\end{pmatrix} \hspace{0.75cm}
	\hat e_y &= \begin{pmatrix}
		0 & 0 & 0 \\
		0 & 0 & 1 \\
		0 & 1 & 0
		\end{pmatrix} \\[3mm]
	\hat e_b &= \begin{pmatrix}
		1 & 0 & 0 \\
		0 & 1 & 0 \\
		0 & 0 & 0
		\end{pmatrix} \hspace{0.75cm}
	\hat e_l &= \begin{pmatrix}
		0 & 0 & 0 \\
		0 & 0 & 0 \\
		0 & 0 & 1
		\end{pmatrix}.
	\end{aligned}
	\end{equation}
The observation of vector or scalar modes would be in direct conflict with general relativity, and so the direct measurement of gravitational-wave polarizations offers a promising avenue by which to test theories of gravity \cite{Will2014}.

Recently, the Advanced LIGO-Virgo network has succeeded in making the first direct statement about the polarization of gravitational waves.
The gravitational-wave signal GW170814, observed by both the Advanced LIGO and Virgo detectors, significantly favored a model assuming pure tensor polarization over models with pure vector or scalar polarizations \cite{GW170814,GW170814_polarization}.
In general, however, the ability of the Advanced LIGO-Virgo network to study the polarization of gravitational-wave transients is limited by several factors.
First, the LIGO-Hanford and LIGO-Livingston detectors are nearly co-oriented, preventing Advanced LIGO from sensitively measuring more than a single polarization mode \cite{GW150914_GR,O1BBH,GW170814,GW170814_polarization}.
Second, at least five detectors are needed to fully characterize the five polarization degrees of freedom accessible to quadrupole detectors.
Quadrupole detectors (those measuring differential arm motion) have degenerate responses to breathing and longitudinal modes, and can therefore measure only a single linear combination of scalar breathing and longitudinal polarizations \cite{Chatziioannou2012,Will2014,2015CQGra..32x3001B,GW170814_polarization}.

Beyond compact binary mergers, another target for Advanced LIGO and Virgo is the stochastic gravitational-wave background.
An astrophysical stochastic background is expected to arise from the population of distant compact binary mergers \cite{Rosado2011,Zhu2011,Wu2012,Zhu2013,StochasticImplications2016,GW170817_stochastic}, core-collapse supernovae \cite{Buonanno2005,Zhu2010,Crocker2015}, and rapidly rotating neutron stars \cite{Rosado2012,Wu2013,Talukder2014}.
In particular, the astrophysical background from compact binary mergers is likely to be detected by LIGO and Virgo at their design sensitivities \cite{GW170817_stochastic}.
A background of cosmological origin may also be present, due to cosmic strings \cite{Olmez2010,PhysRevLett.112.131101}, inflation \cite{Maggiore2000,Giblin2014,Easther2007,Cook2012}, and phase transitions in the early Universe \cite{Maggiore2000,Giblin2014,Caprini2008,Caprini2009,Lopez2015}.

Long duration gravitational-wave sources, like the stochastic background \cite{Nishizawa2009,Nishizawa2010,Nishizawa2013,Callister2017} or persistent signals from rotating neutron stars \cite{Isi2015,Isi2017,CWnonGR}, offer a viable means of searching for nonstandard gravitational-wave polarizations.
Unlike gravitational-wave transients, which sample only a single point on the LIGO/Virgo antenna response patterns, long-duration signals contain information about many points on the antenna patterns.
Long-duration signals, therefore, enable the direct measurement of gravitational-wave polarizations using the current generation of gravitational-wave detectors, without the need for additional detectors or an independent electromagnetic counterpart.
The stochastic background is thus a valuable laboratory for polarization-based tests of general relativity \cite{Callister2017}.

In this Letter, we present the first direct search for vector and scalar polarizations in the stochastic gravitational-wave background.
We analyze data recorded during Advanced LIGO's first observing run (O1), which has previously been searched for both isotropic and anisotropic backgrounds of standard tensor polarizations \cite{IsotropicO1,DirectionalO1}.
First, we describe the O1 data set and its initial processing.
We then discuss the stochastic analysis, including the construction of Bayesian odds that indicate the nondetection of a generically polarized stochastic background in our data.
Finally, we present upper limits on the joint contributions of tensor, vector, and scalar polarizations to the stochastic gravitational-wave background.
Additional details and results are presented in the Supplemental Material, available online.


\textit{Data.} --
We search Advanced LIGO data for evidence of a stochastic background, analyzing data recorded between September 18, 2015 15:00 UTC and January 12, 2016 16:00 UTC during LIGO's O1 observing run.
We do not include several days of O1 data recorded prior to September 18, but this has negligible impact on our results.
We exclude times containing the binary black hole signals GW150914 and GW151226, as well as the signal candidate LVT151012.

The initial data processing proceeds as in previous analyses \cite{IsotropicS6,IsotropicO1}.
Time-domain strain measurements from the LIGO-Hanford and LIGO-Livingston detectors are down-sampled from 16384 Hz to 4096 Hz and divided into half-overlapping 192 s segments.
Each time segment is then Hann-windowed, Fourier transformed, and high-pass filtered using a 16th order Butterworth filter with a knee frequency of 11 Hz.
Finally, the strain data are coarse-grained to a frequency resolution of $0.03125$ Hz and restricted to a frequency band from 20--1726 Hz.
Within each segment, we compute the LIGO-Hanford and LIGO-Livingston strain auto-power spectral densities using Welch's method \cite{Welch}.

Standard data quality cuts are performed in both the time and frequency domains to mitigate the effects of non-Gaussian instrumental and environmental noise \cite{IsotropicO1,IsotropicO1Supplement,DirectionalO1}.
In the time domain, \blue{35}\% of data is discarded due to nonstationary detector noise, leaving \blue{29.85} days of coincident observing time.
In the frequency domain, an additional \blue{21}\% of data is discarded to remove correlated narrow-band features between LIGO-Hanford and LIGO-Livingston \cite{IsotropicO1,IsotropicO1Supplement,DirectionalO1}.
These narrow-band correlations are due to a variety of sources, including injected calibration signals, power mains, and GPS timing systems.
To estimate possible contamination due to terrestrial Schumann resonances \cite{Thrane2013,Thrane2014,Himemoto2017}, we additionally monitored coherences between magnetometers installed at both detectors.
Schumann resonances were found to contribute negligibly to the stochastic measurement \cite{IsotropicO1,IsotropicO1Supplement}.

We assume conservative \blue{4.8\%} and \blue{5.4\%} calibration uncertainties on the strain amplitude measured by LIGO-Hanford and LIGO-Livingston, respectively \cite{Cahillane2017}.
Phase calibration is a much smaller source of uncertainty and is therefore neglected \cite{IsotropicO1,Whelan2014}.
All results below are obtained after marginalization over amplitude uncertainties; see the Supplemental Material for details.


\textit{Method.} --
To search for a generically polarized stochastic background, we will apply the methodology presented in Ref. \cite{Callister2017}.
This method is summarized below, and additional details are discussed in the Supplemental Material.

The stochastic background may be detected in the form of a correlated signal between pairs of gravitational-wave detectors.
We will assume that the stochastic background is stationary, isotropic, and Gaussian.
For simplicity, we also assume that the background is uncorrelated between polarization modes.
Finally, we assume that the tensor and vector contributions to the background are individually unpolarized (with equal contributions, for instance, from the tensor plus and cross modes).
Certain theories may violate one or more of these assumptions.
For example, the stochastic background is unlikely to remain strictly unpolarized in the presence of gravitational-wave birefringence, as in Chern-Simons gravity \cite{PhysRevD.68.104012,2008PhRvD..78f6005A,2009PhR...480....1A}, while theories violating Lorentz invariance may yield a departure from isotropy \cite{2004PhRvD..69j5009K,2016PhLB..757..510K}.
The violation of one or more of our assumptions would likely reduce our search's sensitivity to the stochastic background.

Given the above assumptions, the expected cross-correlation between two detectors in the presence of a stochastic background is of the form \cite{Allen1999,Nishizawa2009,Nishizawa2010,Nishizawa2013}
	\begin{equation}
	\label{crosscorr}
	\langle \tilde s_1(f) \tilde s^*_2(f') \rangle = \frac{1}{2}\delta(f-f') \sum_A \Gamma_A(f) S_h^A(f).
	\end{equation}
Here, $S_h^A(f)$ is the one-sided gravitational-wave strain power spectral density of the net tensor ($A=T$), vector ($V$), and scalar ($S$) contributions to the stochastic background.
The detectors' geometry is encoded in the overlap reduction functions $\Gamma_A(f)$, defined \cite{Nishizawa2009,Allen1999,Christensen1992,Callister2017}
	\begin{equation}
	\Gamma_A(f) = \frac{1}{8\pi} \sum_{a\in A} \int d\hat n\, F_1^a(\hat n) F_2^a(\hat n)\,
		e^{2\pi i f \hat n \cdot \Delta x/c}.
	\end{equation}
$F^a_I(\hat n)$ is the antenna response function of detector $I$ to signals of polarization $a$, $\Delta x$ is the separation vector between detectors, and $c$ is the speed of light.
The integral is taken over all sky directions $\hat n$.

\begin{figure}
\centering
\includegraphics[width=0.48\textwidth]{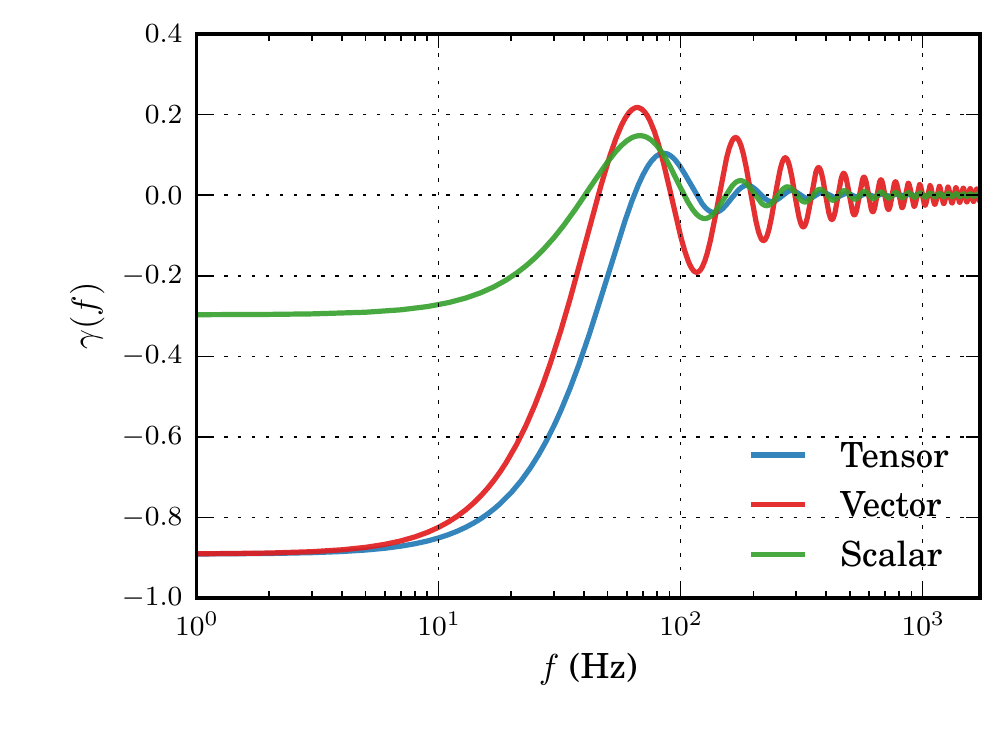}
\caption{
Overlap reduction functions representing the Advanced LIGO network's sensitivity to stochastic backgrounds of tensor (blue), vector (red), and scalar-polarized (green) gravitational waves.
}
\label{orfFigure}
\end{figure}

We will work not directly with $\Gamma_A(f)$, but rather with the \textit{normalized} overlap reduction functions $\gamma_A(f)\propto \Gamma_A(f)/\Gamma_0$, where the constant $\Gamma_0$ is chosen such that $\gamma_T(f) = 1$ for co-located and co-oriented detectors.
For Advanced LIGO, $\Gamma_0 = 1/5$, but in general its value will vary for other experiments like LISA and pulsar timing arrays \cite{Romano2016}.
The normalized overlap reduction functions for LIGO's Hanford-Livingston baseline are shown in Fig. \ref{orfFigure}.
Because tensor, vector, and scalar modes each have distinct overlap reduction functions, the shape of a measured cross-correlation spectrum [Eq. \eqref{crosscorr}] will reflect the polarization content of the stochastic background \cite{Nishizawa2009,Callister2017}.
Of the three curves in Fig. \ref{orfFigure}, the scalar overlap reduction function is smallest in magnitude.
This reflects the fact that the Advanced LIGO detectors have weaker geometrical responses to scalar-polarized gravitational waves than to tensor- and vector-polarized signals.

Conventionally, gravitational-wave backgrounds are parameterized by their energy-density spectra \cite{Allen1999,Romano2016}
	\begin{equation}
	\label{OmegaEq}
	\Omega(f) = \frac{1}{\rho_c} \frac{d\rho_\textsc{gw}}{d\ln f},
	\end{equation}
where $d\rho_\textsc{gw}$ is the energy density in gravitational waves per logarithmic frequency interval $d\ln f$.
We normalize Eq. \eqref{OmegaEq} by $\rho_c = 3 H_0^2 c^2/8\pi G$, the closure energy density of the Universe.
Here, $G$ is Newton's constant and $H_0$ is the Hubble constant; we take $H_0 = 68\,\text{km}\,\text{s}^{-1}\,\text{Mpc}^{-1}$ \cite{Planck}.
The precise relationship between $\Omega(f)$ and $S_h(f)$ is theory dependent.
Under any theory obeying Isaacson's formula for the stress-energy of gravitational waves \cite{PhysRev.166.1272}, the energy-density spectrum is related to $S_h(f)$ by \cite{Allen1999,Callister2017,IsiStein2017}
	\begin{equation}
	\label{energyDensity}
	\Omega(f) = \frac{2\pi^2}{3 H_0^2} f^3 S_h(f).
	\end{equation}
Equation \eqref{energyDensity} does not hold in general, however \cite{IsiStein2017}.
For ease of comparison with previous studies, we will instead take Eq. \eqref{energyDensity} as the \textit{definition} of the canonical energy-density spectra $\Omega^A(f)$.
The canonical energy-density spectra can be directly identified with true energy densities under any theory obeying Isaacson's formula.
For other theories, $\Omega^A(f)$ can instead be understood simply as a function of the detector-frame observable $S_h^A(f)$.

Within each 192 s time segment (indexed by $i$), we form an estimator of the visible cross power between LIGO-Hanford and LIGO-Livingston:
	\begin{equation}
	\hat C_i(f) = \frac{1}{\Delta T}\frac{20\pi^2}{3 H_0^2} f^3 \tilde s^*_{1,i}(f) \tilde s_{2,i}(f),
	\label{C}
	\end{equation}
 normalized such that the estimator's mean and variance are \cite{Callister2017}
	\begin{equation}
	\label{meanC}
 	\langle \hat C_i(f) \rangle = \sum_A \gamma_A(f) \Omega^A(f)
	\end{equation}
and
	\begin{equation}
	\label{sigma}
	\sigma^2_i(f) = \frac{1}{2\Delta T df} \left(\frac{10\pi^2}{3 H_0^2}\right)^2 f^6 P_{1,i}(f) P_{2,i}(f),
	\end{equation}
respectively.
Within Eqs. \eqref{C} and \eqref{sigma}, $\Delta T$ is the segment duration, $df$ the frequency bin width, and $P_{I,i}(f)$ is the one-sided auto-power spectral density of detector $I$ in time segment $i$, defined by
	\begin{equation}
	\langle \tilde s_{I,i}^*(f) \tilde s_{I,i}(f') \rangle = \frac{1}{2} \delta(f-f') P_{I,i}(f).
	\end{equation}
The normalization of $\hat C(f)$ is chosen such that the contribution from each polarization appears symmetrically in Eq.~\eqref{meanC}; this choice differs by a factor of $\gamma_T(f)$ from the point estimate $\hat Y(f)$ typically used in stochastic analyses \cite{IsotropicS6,IsotropicO1,Callister2017}.
Finally, the cross-power estimators from each segment are optimally combined via a weighted sum to form a single cross-power spectrum for the O1 observing run,
	\begin{equation}
	\hat C(f) = \frac{\sum_i \hat C_i(f) \sigma^{-2}_i(f)}{\sum_i \sigma^{-2}_i(f)},
	\label{finalC}
	\end{equation}
with the corresponding variance
	\begin{equation}
	\sigma^{-2}(f) = \sum_i \sigma^{-2}_i(f).
	\label{finalSigma}
	\end{equation}
Note that, unlike transient gravitational-wave searches, searches for the stochastic background are well described by Gaussian statistics due to the large number of time segments contributing to the final cross-power spectrum \cite{Meacher2015}.

Given the measured cross-power spectrum $\hat C(f)$, we compute Bayesian evidence for various hypotheses describing the presence and polarization of a possible stochastic signal within our data.
Evidences are computed using \texttt{PyMultiNest} \cite{Buchner2014}, a Python interface to the nested sampling code \texttt{MultiNest} \cite{Feroz2008,Feroz2009,Feroz2013,Skilling2004,Skilling2006}.
We consider several different hypotheses:
	\begin{itemize}
	\item Gaussian noise (N): No stochastic signal is present in our data, and the measured cross power is due entirely to Gaussian noise.
	\item Signal (SIG): A stochastic background of any polarization(s) is present.
	\item Tensor-polarized (GR): The data contains a purely tensor-polarized stochastic signal, consistent with general relativity.
	\item Nonstandard polarizations (NGR): The data contains a stochastic signal with vector and/or scalar contributions.
	\end{itemize}
These evidences are combined to form two Bayesian odds \cite{Callister2017}:
(1) Odds $\OddsSN$ for the presence of a stochastic signal relative to pure noise, and
(2) odds $\OddsGR$ for the presence of nonstandard polarizations versus ordinary tensor modes.
$\OddsSN$ quantifies evidence for the \textit{detection} of a generically polarized stochastic background, and generally depends only on a background's total power, not its polarization content.
$\OddsGR$ indicates if the background's polarization is inconsistent with general relativity.
In particular, the sensitivity of $\OddsGR$ to nonstandard polarizations is not significantly affected by the strength of any tensor polarization which may also be present \cite{Callister2017}.
See the Supplemental Material for further details about our hypotheses and odds ratio construction, including the priors placed on these hypotheses and their parameters.


\begin{figure*}
\centering
\includegraphics[width=0.98\textwidth]{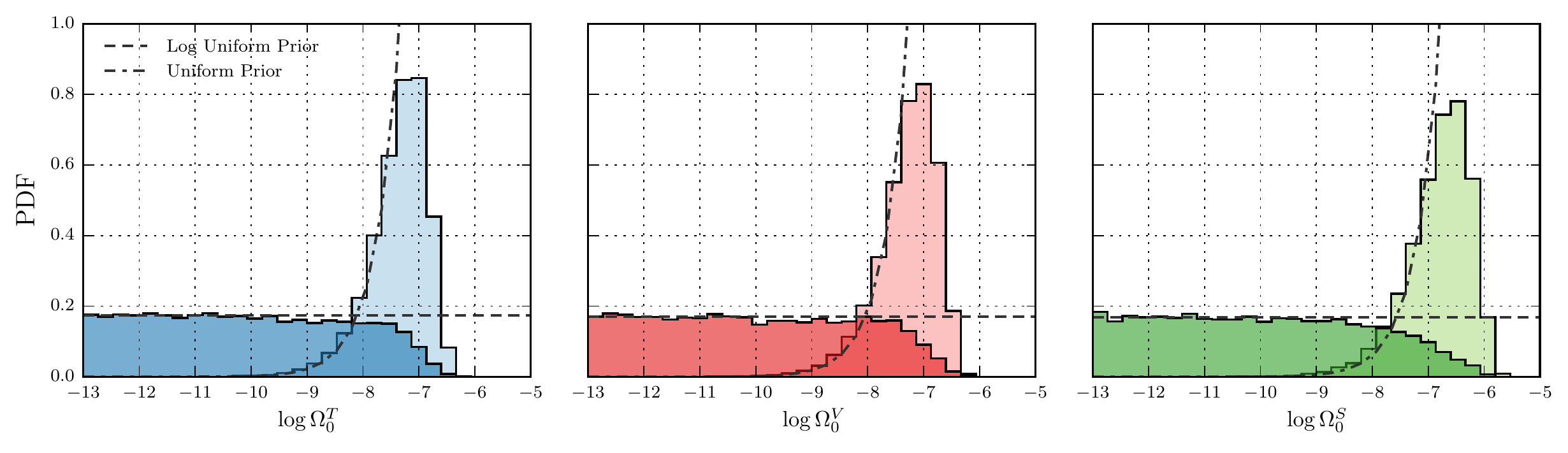}
\caption{
Posteriors on the tensor (left), vector (center), and scalar (right) stochastic background amplitudes at reference frequency $f_0=25$ Hz.
Within each subplot, dark posteriors show results obtained assuming log-uniform priors (dashed curves) on $\Omega^A_0$, while light posteriors show results corresponding to uniform amplitude priors (dot-dashed curves).
The prior curves shown here have been renormalized by constant factors to illustrate consistency with the posteriors below our measured upper limits.
These posteriors correspond to the 95\% credible upper limits listed in Table \ref{peTable}.
Relative to the log-uniform priors, the uniform amplitude priors preferentially weight loud stochastic signals and therefore yield more conservative upper limits.
}
\label{posteriors}
\end{figure*}

\textit{Results.} --
Using the cross power measured between LIGO-Hanford and LIGO-Livingston during Advanced LIGO's O1 observing run, we obtain odds $\ln\OddsSN = \resultOddsSN$ between Signal and Gaussian noise hypotheses, indicating a nondetection of the stochastic gravitational-wave background.
Additionally, we find $\ln\OddsGR = \resultOddsGR$, consistent with values expected in the presence of Gaussian noise \cite{Callister2017}.
(We will use $\ln$ and $\log$ to denote base-$e$ and base-$10$ logarithms, respectively.)

Given our nondetection, we place upper limits on the presence of tensor, vector, and scalar contributions to the stochastic background.
To simultaneously constrain the properties of each polarization, we will restrict our analysis to a model assuming the presence of tensor, vector, and scalar-polarized signals
(this is the TVS hypothesis in the notation of the Supplemental Material).
Under this hypothesis, we model the total canonical energy density of the stochastic background as a sum of power laws:
	\begin{equation}
	\label{TVSequation}
	\Omega(f) = \Omega^T_0 \left(\frac{f}{f_0}\right)^{\alpha_T}
		+ \Omega^V_0 \left(\frac{f}{f_0}\right)^{\alpha_V}
		+ \Omega^S_0 \left(\frac{f}{f_0}\right)^{\alpha_S}.
	\end{equation}
Here, $\Omega^A_0$ is the amplitude of polarization $A$ at a reference frequency $f_0$, and $\alpha_A$ is the corresponding spectral index.
We take $f_0=25$ Hz \cite{IsotropicO1}.
Standard tensor-polarized stochastic backgrounds are predicted to be well described by power laws in the Advanced LIGO band.
The expected astrophysical background from compact binary mergers, for instance, is well modeled by a power law with $\alpha_T = 2/3$ \cite{Rosado2011,Zhu2011,Wu2012,Callister2016}.

\begingroup
\squeezetable
\begin{table}
\caption{95\% credible upper limits on the log amplitudes of tensor, vector, and scalar modes in the stochastic background at reference frequency $f_0 = 25$ Hz.
We assume an energy-density spectrum in which all three modes are present, and present limits following marginalization over the spectral index of each component [see Eq. \eqref{TVSequation}].
We show results for two different amplitude priors: a log-uniform prior ($dp/d\log\Omega_0 \propto 1$; top row) and a uniform prior ($dp/d\Omega_0 \propto 1$; bottom row).
Additional parameter estimation results are shown in the Supplemental Material.
}
\label{peTable}
\setlength{\tabcolsep}{2pt}
\renewcommand{\arraystretch}{1.5}
\begin{tabular}{ l | r r r | r r r}
\hline
\hline
Prior & $\log\Omega^{T}_0$ & $\log\Omega^{V}_0$ & $\log\Omega^{S}_0$
	 & $\Omega^T_0$ & $\Omega^V_0$ & $\Omega^S_0$ \\
\hline
Log uniform & $\logOmgTlogUniformTVS$ 
		& $\logOmgVlogUniformTVS$ 
		& $\logOmgSlogUniformTVS$ 
		& $\num{\OmgTlogUniformTVS}$ 
		& $\num{\OmgVlogUniformTVS}$ 
		& $\num{\OmgSlogUniformTVS}$ \\
Uniform & $\logOmgTuniformTVS$ 
		& $\logOmgVuniformTVS$ 
		& $\logOmgSuniformTVS$
		& $\num{\OmgTuniformTVS}$ 
		& $\num{\OmgVuniformTVS}$ 
		& $\num{\OmgSuniformTVS}$ \\
\hline
\hline
\end{tabular}
\end{table}
\endgroup

We will consider two different prior distributions for the background amplitudes: a log-uniform prior between $10^{-13}\leq\Omega^A_0\leq10^{-5}$ and a uniform prior between $0\leq\Omega^A_0\leq10^{-5}$.
The former (log-uniform) corresponds to the prior adopted in Ref. \cite{Callister2017}.
The latter (uniform) implicitly reproduces the maximum likelihood analysis used in previous studies, and is included to allow direct comparison to previous stochastic results  \cite{IsotropicS6,IsotropicO1}.
The upper amplitude bound ($10^{-5}$) is consistent with limits placed by Initial LIGO and Virgo \cite{IsotropicS6}.
In order to be normalizable, the log-uniform prior requires a nonzero lower bound;
although parameter estimation results will depend on the specific choice of lower bound, in general this dependence is weak \cite{Isi2017}.
Our lower bound ($10^{-13}$) is chosen to encompass small energy densities well below the reach of LIGO and Virgo at design sensitivity \cite{IsotropicO1,GW170817_stochastic}.

Following Ref. \cite{Callister2017}, we take our spectral index priors to be $p(\alpha_A)\propto1-|\alpha_A|/\alpha_\textsc{max}$ for $|\alpha_A|\leq\alpha_\textsc{max}$ and $p(\alpha_A)=0$ elsewhere.
This prior preferentially weights flat energy-density spectra, penalizing spectra which are more steeply positively or negatively sloped in the Advanced LIGO band.
We conservatively choose $\alpha_\textsc{max}=8$, allowing for energy-density spectra significantly steeper than backgrounds predicted from known astrophysical sources (like compact binary mergers).

We perform parameter estimation using posterior samples obtained by \texttt{PyMultiNest}.
Figure \ref{posteriors} shows posteriors on the tensor, vector, and scalar background amplitudes, under each choice of amplitude prior.
The dashed and dot-dashed curves are proportional to the log-uniform and uniform amplitude priors, respectively;
each prior curve has been renormalized by a constant factor to illustrate consistency between our priors and posteriors at small $\Omega^A_0$.
We can now place upper limits on the amplitude of each component at $f_0=25$ Hz.
The 95\% credible upper limits on the amplitude of each polarization are listed in Table \ref{peTable} for each choice of prior (for convenience, we list limits in terms of both $\log\Omega^A_0$ and $\Omega^A_0$).
As no signal was detected, our posteriors on the spectral indices $\alpha_A$ are dominated by our prior.
Full parameter estimation results, including posteriors on $\alpha_A$, are given in the Supplemental Material.

Care should be taken when comparing these upper limits to those obtained in previous analyses (e.g., Table I of Ref. \cite{IsotropicO1}).
Three important distinctions should be kept in mind.
First, the amplitude posteriors in Fig.~\ref{posteriors} (and hence the limits in Table \ref{peTable}) are obtained after marginalization over spectral index.
Previous analysis, on the other hand, typically assume specific fixed spectral indices or present exclusion curves in the $\Omega^T_0 - \alpha_T$ plane \cite{IsotropicO1}.
Second, Bayesian upper limits may be strongly influenced by one's adopted prior.
Uniform amplitude priors, for instance, preferentially weight larger signals and hence yield larger upper limits, while log-uniform priors support smaller signal amplitudes, giving tighter limits.
Finally, our results are obtained under a specific signal hypothesis allowing simultaneously for tensor, vector, and scalar polarizations.
These limits are not generically identical to those that would be obtained if we allowed for tensor modes alone.
In the Supplemental Material, we have tabulated upper limits under a variety of signal hypotheses allowing for each unique combination of gravitational-wave polarizations (our results, though, do not vary considerably between hypotheses).
We have additionally verified that, under the GR (tensor-only) hypothesis with delta-function priors on the background's spectral index, we recover upper limits identical to results previously published in Ref. \cite{IsotropicO1}.


\textit{Conclusion.} --
The direct measurement of gravitational-wave polarizations may open the door to powerful new tests of gravity.
Such measurements largely depend only on the geometry of a gravitational wave's strain and its direction of propagation, not on the details of any specific theory of gravity.
Recently, the Advanced LIGO-Virgo observation of the binary black hole merger GW170814 has enabled the first direct study of gravitational-wave polarizations \cite{GW170814,GW170814_polarization}.
While LIGO and Virgo are limited in their ability to discern the polarization of gravitational-wave transients, the future construction of additional detectors, like KAGRA \cite{KAGRA,KAGRA2} and LIGO-India \cite{LIGOIndia}, will help to break existing degeneracies and allow for increasingly precise polarization measurements.

Long-duration signals offer further opportunities to study gravitational-wave polarizations.
Detections of continuous sources like rotating neutron stars \cite{Isi2017,CWnonGR} and the stochastic background \cite{Callister2017} will offer the ability to directly measure and/or constrain gravitational-wave polarizations, even in the absence of additional detectors.
In this Letter, we have conducted a search for a generically polarized stochastic background of gravitational waves using data from Advanced LIGO's O1 observing run.
Although we find no evidence for the presence of a background (of any polarization), we have succeeded in placing the first direct upper limits (listed in Table \ref{peTable}) on the contributions of vector and scalar modes to the stochastic background.


\vspace{0.1cm}
\input{Acknowledgements}

\bibliography{References}

\clearpage
\onecolumngrid
\begin{center}
{\large\textbf{Supplement To: A Search for Tensor, Vector, and Scalar Polarizations in the Stochastic Gravitational-Wave Background}} \\[0.5cm]
Abbott, B. P. \textit{et al.} \\
(The LIGO Scientific Collaboration and Virgo Collaboration)
\end{center}
\vspace{0.55cm}
\twocolumngrid

\input{Supplement}

\end{document}

%% file: BayesFactors.txt
\newcommand{\resultOddsSN}{-0.53}
\newcommand{\resultOddsGR}{-0.25}
\newcommand{\lnBayesTvsN}{-0.33}
\newcommand{\lnBayesVvsN}{-0.33}
\newcommand{\lnBayesSvsN}{-0.31}
\newcommand{\lnBayesTVvsN}{-0.66}
\newcommand{\lnBayesTSvsN}{-0.65}
\newcommand{\lnBayesVSvsN}{-0.65}
\newcommand{\lnBayesTVSvsN}{-0.99}

%% file: peResults.txt
\newcommand{\logOmgTlogUniformT}{-7.21}
\newcommand{\OmgTlogUniformT}{6.23e-08}
\newcommand{\alphaTlogUniformTmed}{-0.4}
\newcommand{\alphaTlogUniformTlower}{5.9}
\newcommand{\alphaTlogUniformTupper}{5.5}

\newcommand{\logOmgVlogUniformV}{-7.16}
\newcommand{\OmgVlogUniformV}{6.96e-08}
\newcommand{\alphaVlogUniformVmed}{-0.4}
\newcommand{\alphaVlogUniformVlower}{5.8}
\newcommand{\alphaVlogUniformVupper}{4.9}

\newcommand{\logOmgSlogUniformS}{-6.86}
\newcommand{\OmgSlogUniformS}{1.38e-07}
\newcommand{\alphaSlogUniformSmed}{-0.5}
\newcommand{\alphaSlogUniformSlower}{5.8}
\newcommand{\alphaSlogUniformSupper}{5.4}

\newcommand{\logOmgTlogUniformTV}{-7.25}
\newcommand{\logOmgVlogUniformTV}{-7.16}
\newcommand{\OmgTlogUniformTV}{5.56e-08}
\newcommand{\OmgVlogUniformTV}{6.95e-08}
\newcommand{\alphaTlogUniformTVmed}{-0.4}
\newcommand{\alphaTlogUniformTVlower}{5.9}
\newcommand{\alphaTlogUniformTVupper}{5.6}
\newcommand{\alphaVlogUniformTVmed}{-0.4}
\newcommand{\alphaVlogUniformTVlower}{5.9}
\newcommand{\alphaVlogUniformTVupper}{4.8}

\newcommand{\logOmgTlogUniformTS}{-7.25}
\newcommand{\logOmgSlogUniformTS}{-6.93}
\newcommand{\OmgTlogUniformTS}{5.68e-08}
\newcommand{\OmgSlogUniformTS}{1.18e-07}
\newcommand{\alphaTlogUniformTSmed}{-0.4}
\newcommand{\alphaTlogUniformTSlower}{5.9}
\newcommand{\alphaTlogUniformTSupper}{5.5}
\newcommand{\alphaSlogUniformTSmed}{-0.5}
\newcommand{\alphaSlogUniformTSlower}{5.8}
\newcommand{\alphaSlogUniformTSupper}{5.4}

\newcommand{\logOmgVlogUniformVS}{-7.17}
\newcommand{\logOmgSlogUniformVS}{-6.90}
\newcommand{\OmgVlogUniformVS}{6.74e-08}
\newcommand{\OmgSlogUniformVS}{1.25e-07}
\newcommand{\alphaVlogUniformVSmed}{-0.4}
\newcommand{\alphaVlogUniformVSlower}{5.9}
\newcommand{\alphaVlogUniformVSupper}{5.0}
\newcommand{\alphaSlogUniformVSmed}{-0.5}
\newcommand{\alphaSlogUniformVSlower}{5.8}
\newcommand{\alphaSlogUniformVSupper}{5.4}

\newcommand{\logOmgTlogUniformTVS}{-7.25}
\newcommand{\logOmgVlogUniformTVS}{-7.20}
\newcommand{\logOmgSlogUniformTVS}{-6.96}
\newcommand{\OmgTlogUniformTVS}{5.58e-08}
\newcommand{\OmgVlogUniformTVS}{6.35e-08}
\newcommand{\OmgSlogUniformTVS}{1.08e-07}
\newcommand{\alphaTlogUniformTVSmed}{-0.4}
\newcommand{\alphaTlogUniformTVSlower}{5.9}
\newcommand{\alphaTlogUniformTVSupper}{5.5}
\newcommand{\alphaVlogUniformTVSmed}{-0.4}
\newcommand{\alphaVlogUniformTVSlower}{5.9}
\newcommand{\alphaVlogUniformTVSupper}{4.8}
\newcommand{\alphaSlogUniformTVSmed}{-0.6}
\newcommand{\alphaSlogUniformTVSlower}{5.8}
\newcommand{\alphaSlogUniformTVSupper}{5.4}

\newcommand{\logOmgTuniformT}{-6.60}
\newcommand{\OmgTuniformT}{2.53e-07}
\newcommand{\alphaTuniformTmed}{-0.4}
\newcommand{\alphaTuniformTlower}{4.7}
\newcommand{\alphaTuniformTupper}{3.8}

\newcommand{\logOmgVuniformV}{-6.48}
\newcommand{\OmgVuniformV}{3.29e-07}
\newcommand{\alphaVuniformVmed}{-1.5}
\newcommand{\alphaVuniformVlower}{4.9}
\newcommand{\alphaVuniformVupper}{3.3}

\newcommand{\logOmgSuniformS}{-5.97}
\newcommand{\OmgSuniformS}{1.08e-06}
\newcommand{\alphaSuniformSmed}{-2.6}
\newcommand{\alphaSuniformSlower}{4.0}
\newcommand{\alphaSuniformSupper}{4.1}

\newcommand{\logOmgTuniformTV}{-6.66}
\newcommand{\logOmgVuniformTV}{-6.55}
\newcommand{\OmgTuniformTV}{2.18e-07}
\newcommand{\OmgVuniformTV}{2.85e-07}
\newcommand{\alphaTuniformTVmed}{-1.8}
\newcommand{\alphaTuniformTVlower}{4.7}
\newcommand{\alphaTuniformTVupper}{3.9}
\newcommand{\alphaVuniformTVmed}{-1.7}
\newcommand{\alphaVuniformTVlower}{4.9}
\newcommand{\alphaVuniformTVupper}{3.6}

\newcommand{\logOmgTuniformTS}{-6.65}
\newcommand{\logOmgSuniformTS}{-6.03}
\newcommand{\OmgTuniformTS}{2.23e-07}
\newcommand{\OmgSuniformTS}{9.28e-07}
\newcommand{\alphaTuniformTSmed}{-1.7}
\newcommand{\alphaTuniformTSlower}{4.7}
\newcommand{\alphaTuniformTSupper}{3.8}
\newcommand{\alphaSuniformTSmed}{-2.6}
\newcommand{\alphaSuniformTSlower}{4.1}
\newcommand{\alphaSuniformTSupper}{4.0}

\newcommand{\logOmgVuniformVS}{-6.53}
\newcommand{\logOmgSuniformVS}{-6.04}
\newcommand{\OmgVuniformVS}{2.96e-07}
\newcommand{\OmgSuniformVS}{9.21e-07}
\newcommand{\alphaVuniformVSmed}{-1.7}
\newcommand{\alphaVuniformVSlower}{4.7}
\newcommand{\alphaVuniformVSupper}{3.6}
\newcommand{\alphaSuniformVSmed}{-2.8}
\newcommand{\alphaSuniformVSlower}{4.0}
\newcommand{\alphaSuniformVSupper}{4.1}

\newcommand{\logOmgTuniformTVS}{-6.70}
\newcommand{\logOmgVuniformTVS}{-6.59}
\newcommand{\logOmgSuniformTVS}{-6.07}
\newcommand{\OmgTuniformTVS}{2.02e-07}
\newcommand{\OmgVuniformTVS}{2.54e-07}
\newcommand{\OmgSuniformTVS}{8.44e-07}
\newcommand{\alphaTuniformTVSmed}{-1.7}
\newcommand{\alphaTuniformTVSlower}{4.8}
\newcommand{\alphaTuniformTVSupper}{3.8}
\newcommand{\alphaVuniformTVSmed}{-1.7}
\newcommand{\alphaVuniformTVSlower}{4.7}
\newcommand{\alphaVuniformTVSupper}{3.7}
\newcommand{\alphaSuniformTVSmed}{-2.6}
\newcommand{\alphaSuniformTVSlower}{4.1}
\newcommand{\alphaSuniformTVSupper}{4.0}

%% file: LVC_Aug2017_StochasticTensorVectorScalar-prd.tex
\author{%
B.~P.~Abbott,$^{1}$  
R.~Abbott,$^{1}$  
T.~D.~Abbott,$^{2}$  
F.~Acernese,$^{3,4}$ 
K.~Ackley,$^{5,6}$  
C.~Adams,$^{7}$  
T.~Adams,$^{8}$ 
P.~Addesso,$^{9}$  
R.~X.~Adhikari,$^{1}$  
V.~B.~Adya,$^{10}$
% ...
\\
\medskip
(LIGO Scientific Collaboration and Virgo Collaboration) 
\\
\medskip
{${}^{\ast}$Deceased, February 2017. }%
{${}^{\dag}$Deceased, November 2017. }%
{${}^{\ddag}$Deceased, December 2016. }%
}\noaffiliation
\affiliation {LIGO, California Institute of Technology, Pasadena, CA 91125, USA }
\affiliation {Louisiana State University, Baton Rouge, LA 70803, USA }
\affiliation {Universit\`a di Salerno, Fisciano, I-84084 Salerno, Italy }
\affiliation {INFN, Sezione di Napoli, Complesso Universitario di Monte S.Angelo, I-80126 Napoli, Italy }
\affiliation {University of Florida, Gainesville, FL 32611, USA }
\affiliation {OzGrav, School of Physics \& Astronomy, Monash University, Clayton 3800, Victoria, Australia }
\affiliation {LIGO Livingston Observatory, Livingston, LA 70754, USA }
\affiliation {Laboratoire d'Annecy-le-Vieux de Physique des Particules (LAPP), Universit\'e Savoie Mont Blanc, CNRS/IN2P3, F-74941 Annecy, France }
\affiliation {University of Sannio at Benevento, I-82100 Benevento, Italy and INFN, Sezione di Napoli, I-80100 Napoli, Italy }
\affiliation {Max Planck Institute for Gravitational Physics (Albert Einstein Institute), D-30167 Hannover, Germany }
\affiliation {The University of Mississippi, University, MS 38677, USA }
\affiliation {NCSA, University of Illinois at Urbana-Champaign, Urbana, IL 61801, USA }
\affiliation {University of Cambridge, Cambridge CB2 1TN, United Kingdom }
\affiliation {Nikhef, Science Park, 1098 XG Amsterdam, The Netherlands }
\affiliation {LIGO, Massachusetts Institute of Technology, Cambridge, MA 02139, USA }
\affiliation {Instituto Nacional de Pesquisas Espaciais, 12227-010 S\~{a}o Jos\'{e} dos Campos, S\~{a}o Paulo, Brazil }
\affiliation {Gran Sasso Science Institute (GSSI), I-67100 L'Aquila, Italy }
\affiliation {INFN, Laboratori Nazionali del Gran Sasso, I-67100 Assergi, Italy }
\affiliation {Inter-University Centre for Astronomy and Astrophysics, Pune 411007, India }
\affiliation {International Centre for Theoretical Sciences, Tata Institute of Fundamental Research, Bengaluru 560089, India }
\affiliation {University of Wisconsin-Milwaukee, Milwaukee, WI 53201, USA }
\affiliation {Leibniz Universit\"at Hannover, D-30167 Hannover, Germany }
\affiliation {Universit\`a di Pisa, I-56127 Pisa, Italy }
\affiliation {INFN, Sezione di Pisa, I-56127 Pisa, Italy }
\affiliation {OzGrav, Australian National University, Canberra, Australian Capital Territory 0200, Australia }
\affiliation {Laboratoire des Mat\'eriaux Avanc\'es (LMA), CNRS/IN2P3, F-69622 Villeurbanne, France }
\affiliation {SUPA, University of the West of Scotland, Paisley PA1 2BE, United Kingdom }
\affiliation {LAL, Univ. Paris-Sud, CNRS/IN2P3, Universit\'e Paris-Saclay, F-91898 Orsay, France }
\affiliation {California State University Fullerton, Fullerton, CA 92831, USA }
\affiliation {European Gravitational Observatory (EGO), I-56021 Cascina, Pisa, Italy }
\affiliation {Universit\`a di Roma Tor Vergata, I-00133 Roma, Italy }
\affiliation {INFN, Sezione di Roma Tor Vergata, I-00133 Roma, Italy }
\affiliation {Universit\"at Hamburg, D-22761 Hamburg, Germany }
\affiliation {INFN, Sezione di Roma, I-00185 Roma, Italy }
\affiliation {Cardiff University, Cardiff CF24 3AA, United Kingdom }
\affiliation {Embry-Riddle Aeronautical University, Prescott, AZ 86301, USA }
\affiliation {Max Planck Institute for Gravitational Physics (Albert Einstein Institute), D-14476 Potsdam-Golm, Germany }
\affiliation {APC, AstroParticule et Cosmologie, Universit\'e Paris Diderot, CNRS/IN2P3, CEA/Irfu, Observatoire de Paris, Sorbonne Paris Cit\'e, F-75205 Paris Cedex 13, France }
\affiliation {Korea Institute of Science and Technology Information, Daejeon 34141, Korea }
\affiliation {West Virginia University, Morgantown, WV 26506, USA }
\affiliation {Universit\`a di Perugia, I-06123 Perugia, Italy }
\affiliation {INFN, Sezione di Perugia, I-06123 Perugia, Italy }
\affiliation {Syracuse University, Syracuse, NY 13244, USA }
\affiliation {University of Minnesota, Minneapolis, MN 55455, USA }
\affiliation {SUPA, University of Glasgow, Glasgow G12 8QQ, United Kingdom }
\affiliation {LIGO Hanford Observatory, Richland, WA 99352, USA }
\affiliation {Caltech CaRT, Pasadena, CA 91125, USA }
\affiliation {Wigner RCP, RMKI, H-1121 Budapest, Konkoly Thege Mikl\'os \'ut 29-33, Hungary }
\affiliation {Columbia University, New York, NY 10027, USA }
\affiliation {Stanford University, Stanford, CA 94305, USA }
\affiliation {Universit\`a di Camerino, Dipartimento di Fisica, I-62032 Camerino, Italy }
\affiliation {Universit\`a di Padova, Dipartimento di Fisica e Astronomia, I-35131 Padova, Italy }
\affiliation {INFN, Sezione di Padova, I-35131 Padova, Italy }
\affiliation {Institute of Physics, E\"otv\"os University, P\'azm\'any P. s. 1/A, Budapest 1117, Hungary }
\affiliation {Nicolaus Copernicus Astronomical Center, Polish Academy of Sciences, 00-716, Warsaw, Poland }
\affiliation {Rochester Institute of Technology, Rochester, NY 14623, USA }
\affiliation {University of Birmingham, Birmingham B15 2TT, United Kingdom }
\affiliation {INFN, Sezione di Genova, I-16146 Genova, Italy }
\affiliation {RRCAT, Indore MP 452013, India }
\affiliation {Faculty of Physics, Lomonosov Moscow State University, Moscow 119991, Russia }
\affiliation {SUPA, University of Strathclyde, Glasgow G1 1XQ, United Kingdom }
\affiliation {The Pennsylvania State University, University Park, PA 16802, USA }
\affiliation {OzGrav, University of Western Australia, Crawley, Western Australia 6009, Australia }
\affiliation {Department of Astrophysics/IMAPP, Radboud University Nijmegen, P.O. Box 9010, 6500 GL Nijmegen, The Netherlands }
\affiliation {Artemis, Universit\'e C\^ote d'Azur, Observatoire C\^ote d'Azur, CNRS, CS 34229, F-06304 Nice Cedex 4, France }
\affiliation {Institut FOTON, CNRS, Universit\'e de Rennes 1, F-35042 Rennes, France }
\affiliation {Washington State University, Pullman, WA 99164, USA }
\affiliation {University of Oregon, Eugene, OR 97403, USA }
\affiliation {Laboratoire Kastler Brossel, UPMC-Sorbonne Universit\'es, CNRS, ENS-PSL Research University, Coll\`ege de France, F-75005 Paris, France }
\affiliation {Carleton College, Northfield, MN 55057, USA }
\affiliation {OzGrav, University of Adelaide, Adelaide, South Australia 5005, Australia }
\affiliation {Astronomical Observatory Warsaw University, 00-478 Warsaw, Poland }
\affiliation {VU University Amsterdam, 1081 HV Amsterdam, The Netherlands }
\affiliation {University of Maryland, College Park, MD 20742, USA }
\affiliation {School of Physics, Georgia Institute of Technology, Atlanta, GA 30332, USA }
\affiliation {Universit\'e Claude Bernard Lyon 1, F-69622 Villeurbanne, France }
\affiliation {Universit\`a di Napoli `Federico II,' Complesso Universitario di Monte S.Angelo, I-80126 Napoli, Italy }
\affiliation {NASA Goddard Space Flight Center, Greenbelt, MD 20771, USA }
\affiliation {Dipartimento di Fisica, Universit\`a degli Studi di Genova, I-16146 Genova, Italy }
\affiliation {RESCEU, University of Tokyo, Tokyo, 113-0033, Japan. }
\affiliation {Tsinghua University, Beijing 100084, China }
\affiliation {Texas Tech University, Lubbock, TX 79409, USA }
\affiliation {Kenyon College, Gambier, OH 43022, USA }
\affiliation {Departamento de Astronom\'{\i }a y Astrof\'{\i }sica, Universitat de Val\`encia, E-46100 Burjassot, Val\`encia, Spain }
\affiliation {Museo Storico della Fisica e Centro Studi e Ricerche Enrico Fermi, I-00184 Roma, Italy }
\affiliation {National Tsing Hua University, Hsinchu City, 30013 Taiwan, Republic of China }
\affiliation {Charles Sturt University, Wagga Wagga, New South Wales 2678, Australia }
\affiliation {Center for Interdisciplinary Exploration \& Research in Astrophysics (CIERA), Northwestern University, Evanston, IL 60208, USA }
\affiliation {University of Chicago, Chicago, IL 60637, USA }
\affiliation {Pusan National University, Busan 46241, Korea }
\affiliation {The Chinese University of Hong Kong, Shatin, NT, Hong Kong }
\affiliation {INAF, Osservatorio Astronomico di Padova, I-35122 Padova, Italy }
\affiliation {INFN, Trento Institute for Fundamental Physics and Applications, I-38123 Povo, Trento, Italy }
\affiliation {OzGrav, University of Melbourne, Parkville, Victoria 3010, Australia }
\affiliation {Universit\`a di Roma `La Sapienza,' I-00185 Roma, Italy }
\affiliation {Universit\'e Libre de Bruxelles, Brussels 1050, Belgium }
\affiliation {Sonoma State University, Rohnert Park, CA 94928, USA }
\affiliation {Departamento de Matem\'aticas, Universitat de Val\`encia, E-46100 Burjassot, Val\`encia, Spain }
\affiliation {Montana State University, Bozeman, MT 59717, USA }
\affiliation {Universitat de les Illes Balears, IAC3---IEEC, E-07122 Palma de Mallorca, Spain }
\affiliation {The University of Texas Rio Grande Valley, Brownsville, TX 78520, USA }
\affiliation {Bellevue College, Bellevue, WA 98007, USA }
\affiliation {Institute for Plasma Research, Bhat, Gandhinagar 382428, India }
\affiliation {The University of Sheffield, Sheffield S10 2TN, United Kingdom }
\affiliation {Dipartimento di Scienze Matematiche, Fisiche e Informatiche, Universit\`a di Parma, I-43124 Parma, Italy }
\affiliation {INFN, Sezione di Milano Bicocca, Gruppo Collegato di Parma, I-43124 Parma, Italy }
\affiliation {California State University, Los Angeles, 5151 State University Dr, Los Angeles, CA 90032, USA }
\affiliation {Universit\`a di Trento, Dipartimento di Fisica, I-38123 Povo, Trento, Italy }
\affiliation {Montclair State University, Montclair, NJ 07043, USA }
\affiliation {National Astronomical Observatory of Japan, 2-21-1 Osawa, Mitaka, Tokyo 181-8588, Japan }
\affiliation {Canadian Institute for Theoretical Astrophysics, University of Toronto, Toronto, Ontario M5S 3H8, Canada }
\affiliation {Observatori Astron\`omic, Universitat de Val\`encia, E-46980 Paterna, Val\`encia, Spain }
\affiliation {School of Mathematics, University of Edinburgh, Edinburgh EH9 3FD, United Kingdom }
\affiliation {University and Institute of Advanced Research, Koba Institutional Area, Gandhinagar Gujarat 382007, India }
\affiliation {IISER-TVM, CET Campus, Trivandrum Kerala 695016, India }
\affiliation {University of Szeged, D\'om t\'er 9, Szeged 6720, Hungary }
\affiliation {University of Michigan, Ann Arbor, MI 48109, USA }
\affiliation {Tata Institute of Fundamental Research, Mumbai 400005, India }
\affiliation {INAF, Osservatorio Astronomico di Capodimonte, I-80131, Napoli, Italy }
\affiliation {Universit\`a degli Studi di Urbino `Carlo Bo,' I-61029 Urbino, Italy }
\affiliation {INFN, Sezione di Firenze, I-50019 Sesto Fiorentino, Firenze, Italy }
\affiliation {Physik-Institut, University of Zurich, Winterthurerstrasse 190, 8057 Zurich, Switzerland }
\affiliation {American University, Washington, D.C. 20016, USA }
\affiliation {University of Bia{\l }ystok, 15-424 Bia{\l }ystok, Poland }
\affiliation {University of Southampton, Southampton SO17 1BJ, United Kingdom }
\affiliation {University of Washington Bothell, 18115 Campus Way NE, Bothell, WA 98011, USA }
\affiliation {Institute of Applied Physics, Nizhny Novgorod, 603950, Russia }
\affiliation {Korea Astronomy and Space Science Institute, Daejeon 34055, Korea }
\affiliation {Inje University Gimhae, South Gyeongsang 50834, Korea }
\affiliation {National Institute for Mathematical Sciences, Daejeon 34047, Korea }
\affiliation {NCBJ, 05-400 \'Swierk-Otwock, Poland }
\affiliation {Institute of Mathematics, Polish Academy of Sciences, 00656 Warsaw, Poland }
\affiliation {Hillsdale College, Hillsdale, MI 49242, USA }
\affiliation {Hanyang University, Seoul 04763, Korea }
\affiliation {Seoul National University, Seoul 08826, Korea }
\affiliation {NASA Marshall Space Flight Center, Huntsville, AL 35811, USA }
\affiliation {ESPCI, CNRS, F-75005 Paris, France }
\affiliation {Southern University and A\&M College, Baton Rouge, LA 70813, USA }
\affiliation {College of William and Mary, Williamsburg, VA 23187, USA }
\affiliation {Centre Scientifique de Monaco, 8 quai Antoine Ier, MC-98000, Monaco }
\affiliation {Indian Institute of Technology Madras, Chennai 600036, India }
\affiliation {IISER-Kolkata, Mohanpur, West Bengal 741252, India }
\affiliation {Whitman College, 345 Boyer Avenue, Walla Walla, WA 99362 USA }
\affiliation {Indian Institute of Technology Bombay, Powai, Mumbai, Maharashtra 400076, India }
\affiliation {Scuola Normale Superiore, Piazza dei Cavalieri 7, I-56126 Pisa, Italy }
\affiliation {Universit\'e de Lyon, F-69361 Lyon, France }
\affiliation {Hobart and William Smith Colleges, Geneva, NY 14456, USA }
\affiliation {OzGrav, Swinburne University of Technology, Hawthorn VIC 3122, Australia }
\affiliation {Janusz Gil Institute of Astronomy, University of Zielona G\'ora, 65-265 Zielona G\'ora, Poland }
\affiliation {University of Washington, Seattle, WA 98195, USA }
\affiliation {King's College London, University of London, London WC2R 2LS, United Kingdom }
\affiliation {Indian Institute of Technology, Gandhinagar Ahmedabad Gujarat 382424, India }
\affiliation {Indian Institute of Technology Hyderabad, Sangareddy, Khandi, Telangana 502285, India }
\affiliation {International Institute of Physics, Universidade Federal do Rio Grande do Norte, Natal RN 59078-970, Brazil }
\affiliation {Andrews University, Berrien Springs, MI 49104, USA }
\affiliation {Universit\`a di Siena, I-53100 Siena, Italy }
\affiliation {Trinity University, San Antonio, TX 78212, USA }
\affiliation {Van Swinderen Institute for Particle Physics and Gravity, University of Groningen, Nijenborgh 4, 9747 AG Groningen, The Netherlands }
\affiliation {Abilene Christian University, Abilene, TX 79699, USA }
\affiliation {Colorado State University, Fort Collins, CO 80523, USA }

%% file: Acknowledgements.tex
The authors gratefully acknowledge the support of the United States
National Science Foundation (NSF) for the construction and operation of the
LIGO Laboratory and Advanced LIGO as well as the Science and Technology Facilities Council (STFC) of the
United Kingdom, the Max-Planck-Society (MPS), and the State of
Niedersachsen/Germany for support of the construction of Advanced LIGO 
and construction and operation of the GEO600 detector. 
Additional support for Advanced LIGO was provided by the Australian Research Council.
The authors gratefully acknowledge the Italian Istituto Nazionale di Fisica Nucleare (INFN),  
the French Centre National de la Recherche Scientifique (CNRS) and
the Foundation for Fundamental Research on Matter supported by the Netherlands Organisation for Scientific Research, 
for the construction and operation of the Virgo detector
and the creation and support  of the EGO consortium. 
The authors also gratefully acknowledge research support from these agencies as well as by 
the Council of Scientific and Industrial Research of India, 
the Department of Science and Technology, India,
the Science \& Engineering Research Board (SERB), India,
the Ministry of Human Resource Development, India,
the Spanish  Agencia Estatal de Investigaci\'on,
the Vicepresid\`encia i Conselleria d'Innovaci\'o, Recerca i Turisme and the Conselleria d'Educaci\'o i Universitat del Govern de les Illes Balears,
the Conselleria d'Educaci\'o, Investigaci\'o, Cultura i Esport de la Generalitat Valenciana,
the National Science Centre of Poland,
the Swiss National Science Foundation (SNSF),
the Russian Foundation for Basic Research, 
the Russian Science Foundation,
the European Commission,
the European Regional Development Funds (ERDF),
the Royal Society, 
the Scottish Funding Council, 
the Scottish Universities Physics Alliance, 
the Hungarian Scientific Research Fund (OTKA),
the Lyon Institute of Origins (LIO),
the Paris \^{I}le-de-France Region, 
the National Research, Development and Innovation Office Hungary (NKFI), 
the National Research Foundation of Korea,
Industry Canada and the Province of Ontario through the Ministry of Economic Development and Innovation, 
the Natural Science and Engineering Research Council Canada,
the Canadian Institute for Advanced Research,
the Brazilian Ministry of Science, Technology, Innovations, and Communications,
the International Center for Theoretical Physics South American Institute for Fundamental Research (ICTP-SAIFR), 
the Research Grants Council of Hong Kong,
the National Natural Science Foundation of China (NSFC),
the Leverhulme Trust, 
the Research Corporation, 
the Ministry of Science and Technology (MOST), Taiwan
and
the Kavli Foundation.
The authors gratefully acknowledge the support of the NSF, STFC, MPS, INFN, CNRS and the
State of Niedersachsen/Germany for provision of computational resources.

%% file: Supplement.tex
\section{Sensitive Frequency Bands}

Although this search utilizes the full 20-1726 Hz frequency band, different frequency sub-bands contribute variously to our overall search sensitivity.
To illustrate this, we can investigate the contribution from each frequency bin to a background's optimal signal-to-noise ratio (SNR), given by \cite{Allen1999}
	\begin{equation}
	\label{SNR}
	\text{SNR}^2 = \frac{3 H_0^2}{10\pi^2} 2 T \int_0^\infty \frac{ \left[\sum_A \gamma_A(f)\Omega^A(f)\right]^2}{f^6 P_1(f) P_2(f)} df.
	\end{equation}
Up to additive constants, SNR and $\OddsSN$ are related by $\ln\OddsSN \sim \text{SNR}^2/2$.

\begin{figure}
\centering
\includegraphics[width=0.48\textwidth]{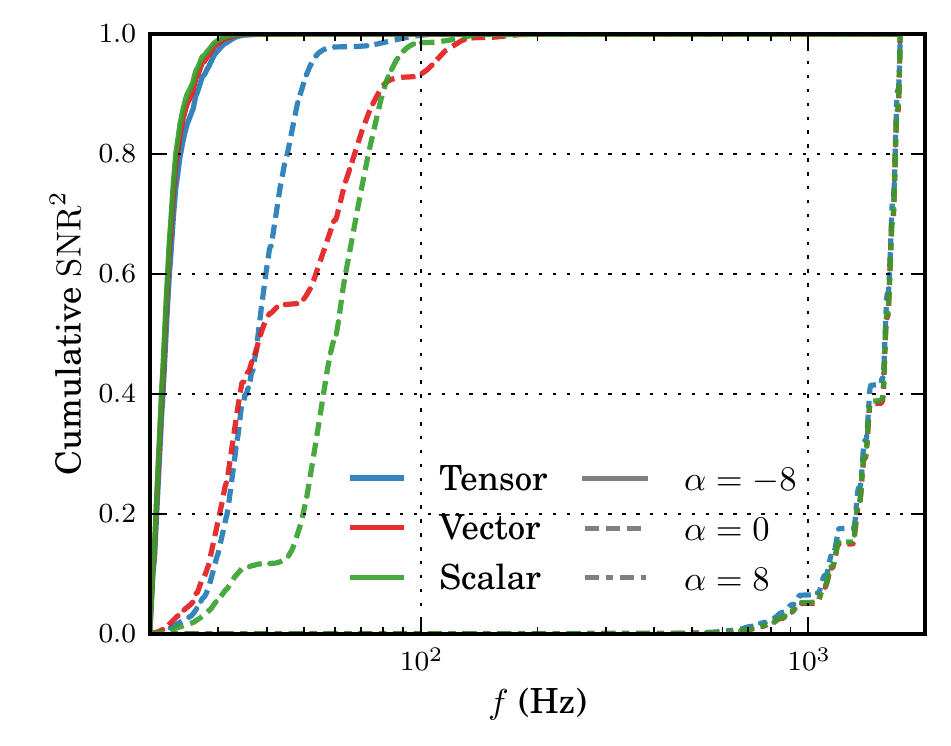}
\caption{
Cumulative squared signal-to-noise ratios as a function of frequency for hypothetical backgrounds of tensor (blue), vector (red), scalar (green) polarizations with spectral indices $\alpha=-8$, $0$, and $8$ (solid, dashed, and dot-dashed, respectively).
The three $\alpha=-8$ curves lie nearly on top of one another, as do the three $\alpha=8$ curves.
The Advanced LIGO network is most sensitive to negatively-sloped backgrounds at low frequencies, while high frequencies contribute the most sensitively to positively-sloped backgrounds.
}
\label{cumulativeSNR}
\end{figure}

Using the measured O1 search sensitivity, Fig. \ref{cumulativeSNR} illustrates the cumulative fraction of the squared-SNR of several representative hypothetical backgrounds, obtained by integrating Eq. \eqref{SNR} from 20 Hz up to a cutoff frequency $f$.
Results are shown for purely tensor- (blue), vector- (red), and scalar-polarized (green) backgrounds, with spectral indices $\alpha=-8$, $0$, and $8$.

As seen in Fig. \ref{cumulativeSNR}, the most sensitive frequency band for a given background is highly dependent on the background's spectral index.
For steeply negatively-sloped backgrounds ($\alpha=-8$), the majority of the measured SNR is obtained at very low frequencies between $\sim20-30$ Hz.
Meanwhile, the $\sim20-100$ Hz band is most sensitive to flat backgrounds, and high frequencies above $\sim700$ Hz are most sensitive to steeply positively-sloped backgrounds.
Although trends are generally independent of polarization, Fig. \ref{cumulativeSNR} does show somewhat different behaviors for tensor, vector, and scalar modes.
These differences are due to the different overlap reduction functions for each polarization sector.

\section{Model Construction}

Here, we briefly summarize the construction of our Signal, Gaussian noise, Non-standard polarization, and Tensor-polarization hypotheses; see Ref. \cite{Callister2017} for further details.

\textit{Gaussian noise}:
We assume that no signal is present and the observed cross-power $\hat C(f)$ is Gaussian distributed about zero with variance given by Eqs. \eqref{sigma} and \eqref{finalSigma}.
Although Advanced LIGO instrumental noise is neither stationary nor Gaussian, searches for the stochastic background are nonetheless well-described by Gaussian statistics due to the large number of time-segments combined to form the final cross-power spectrum $\hat C(f)$ \cite{Meacher2015}.

\textit{Signal}:
The Signal hypothesis is the union of seven sub-hypotheses, which together allow for each unique combination of tensor, vector, and scalar polarizations.
The ``TVS'' sub-hypothesis, for example, assumes the simultaneous presence of all polarization modes, with a canonical energy-density spectrum of the form:
	\begin{equation}
	\Omega_\textsc{tvs}(f) = \Omega^T_0 \left(\frac{f}{f_0}\right)^{\alpha_T}
		+ \Omega^V_0 \left(\frac{f}{f_0}\right)^{\alpha_V}
		+ \Omega^S_0 \left(\frac{f}{f_0}\right)^{\alpha_S}.
	\end{equation}
The ``TS'' sub-hypothesis, meanwhile, assumes only the existence of tensor and scalar modes:
	\begin{equation}
	\Omega_\textsc{ts}(f) = \Omega^T_0 \left(\frac{f}{f_0}\right)^{\alpha_T}
		+ \Omega^S_0 \left(\frac{f}{f_0}\right)^{\alpha_S}.
	\end{equation}
In this fashion, we can construct seven unique sub-hypotheses: \{T,V,S,TV,TS,VS,TVS\}.
The union of these seven possibilities is the Signal hypothesis.

\textit{Non-standard polarization (NGR)} --
Analogous to the Signal hypothesis above, this is the union of the six sub-hypotheses \{V,S,TV,TS,VS,TVS\} containing non-standard polarizations.

\textit{Tensor-polarization (GR)} --
We assume the stochastic background is present and purely-tensor polarized, with the energy-density spectrum
	\begin{equation}
	\Omega_\textsc{gr} = \Omega^T_0 \left(\frac{f}{f_0}\right)^{\alpha_T};
	\end{equation}
this hypothesis is identical the ``T'' signal sub-hypothesis above.

\section{Odds Ratios}

Here, we review the procedure for constructing odds $\OddsSN$ between Signal and Gaussian noise hypotheses, and odds $\OddsGR$ between the NGR and GR hypotheses.
As above, see Ref. \cite{Callister2017} for further details.

The odds between two hypotheses $\mathcal{M}$ and $\mathcal{N}$ is the ratio of posterior probabilities for each hypothesis, given data $d$:
	\begin{equation}
	\begin{aligned}
	\mathcal{O}^\smallmathcal{M}_\smallmathcal{N}
		&= \frac{p(\mathcal{M}|d)}{p(\mathcal{N}|d)}\\
		&= \mathcal{B}^\smallmathcal{M}_\smallmathcal{N} \frac{\pi(\mathcal{M})}{\pi(\mathcal{N})},
	\end{aligned}
	\end{equation}
where $\mathcal{B}^\smallmathcal{M}_\smallmathcal{N}$ is the Bayes factor between the two hypotheses and $\pi(\mathcal{M})$ and $\pi(\mathcal{N})$ are the prior probabilities on $\mathcal{M}$ and $\mathcal{N}$, respectively.
The ratio $\pi(\mathcal{M})/\pi(\mathcal{N})$ is known as the prior odds.

To obtain $\OddsSN$, we first compute the Bayes factor $\mathcal{B}^\smallmathcal{A}_\textsc{n}$ between each Signal sub-hypothesis $\mathcal{A}\in\{\text{T},\text{V},\text{S},...\}$ and the Noise hypothesis.
Because each sub-hypothesis is independent, $\OddsSN$ is then just the sum
	\begin{equation}
	\label{OddsSN}
	\begin{aligned}
	\OddsSN &= \sum_{\smallmathcal{A}\,\in\,\textsc{sig}} \mathcal{O}^\smallmathcal{A}_\textsc{n} \\
		&= \sum_{\smallmathcal{A}\,\in\,\textsc{sig}} \mathcal{B}^\smallmathcal{A}_\textsc{n} \frac{\pi(\mathcal{A}|\text{SIG})\pi(\text{SIG})}{\pi(\text{N})},
	\end{aligned}
	\end{equation}
where we have expanded $\pi(\mathcal{A}) = \pi(\mathcal{A}|\text{SIG})\pi(\text{SIG})$.
We choose equal prior probabilities on the Signal and Noise hypotheses, such that $\pi(\text{SIG})/\pi(\text{N}) = 1$.
Within the Signal hypothesis, we assign equal probabilities to each sub-hypothesis, giving $\pi(\mathcal{A}|\text{SIG}) = 1/7$.

The odds $\OddsGR$ is analogously given by
	\begin{equation}
	\label{OddsGR}
	\OddsGR = \sum_{\smallmathcal{A}\,\in\,\textsc{ngr}} \mathcal{B}^\smallmathcal{A}_\textsc{gr} \frac{\pi(\mathcal{A}|\text{NGR})\pi(\text{NGR})}{\pi(\text{GR})}.
	\end{equation}
We set $\pi(\text{NGR})/\pi(\text{GR}) = 1$ and again choose equal prior probabilities for each sub-hypotheses within NGR, such that $\pi(\mathcal{A}|\text{NGR}) = 1/6$.
	
Our chosen prior odds between hypotheses are necessarily somewhat arbitrary, and different choices will yield different values of $\OddsSN$ and $\OddsGR$.
For completeness, Table \ref{oddsTable} provides the Bayes factors between each signal sub-hypothesis and Gaussian noise.
These Bayes factors allow readers to recompute odds $\OddsSN$ and $\OddsGR$ using different choices of prior odds.

\begin{table}[t!]
\caption{Bayes factors between each signal sub-hypothesis and the Gaussian noise hypothesis, as computed by \texttt{MultiNest}.
These Bayes factors are combined following Eqs. \eqref{OddsSN} and \eqref{OddsGR} to obtain odds $\OddsSN$ between Signal and Gaussian noise hypotheses, and odds $\OddsGR$ between NGR and GR hypotheses.
}
\label{oddsTable}
\setlength{\tabcolsep}{3pt}
\renewcommand{\arraystretch}{1.4}
\begin{tabular}{l | r }
\hline
\hline
Hypothesis & $\ln\mathcal{B}^\mathcal{A}_\textsc{n}$ \\
\hline
T & $\lnBayesTvsN$ \\
V & $\lnBayesVvsN$ \\
S & $\lnBayesSvsN$ \\
TV & $\lnBayesTVvsN$ \\
TS & $\lnBayesTSvsN$ \\
VS & $\lnBayesVSvsN$ \\
TVS & $\lnBayesTVSvsN$ \\
\hline
\hline
\end{tabular}
\end{table}

\section{Calibration Uncertainty}

The strain measured by LIGO-Hanford and LIGO-Livingston is not known perfectly, but is subject to non-zero calibration uncertainty.
For imperfectly calibrated data, the cross-power measurements $\hat C(f)$ are not estimators of $\sum_A \gamma_A(f)\Omega^A(f)$, but rather of $\lambda \sum_A \gamma_A(f)\Omega^A(f)$, where $\lambda$ is some multiplicative factor \cite{Whelan2014}.
Perfect calibration would yield $\lambda=1$, but in general $\lambda$ is unknown.
We include the calibration factor $\lambda$ as an additional parameter in \texttt{MultiNest}, so that the likelihood function becomes
	\begin{equation}
	\begin{aligned}
	\mathcal{L}(&\hat C(f) | \Omega^A_0, \alpha_A, \lambda) \\
		&\propto \prod_f \exp \left[ - \frac{ \left( \hat C(f) - \lambda \sum_A \gamma_A(f) \Omega^A_0 (f/f_0)^{\alpha_A} \right)^2 }{2\sigma^2(f)}\right],
	\end{aligned}
	\end{equation}
with $\hat C(f)$ and $\sigma^2(f)$ given by Eqs. \eqref{finalC} and \eqref{finalSigma}, respectively.

We place a Gaussian prior on $\lambda$, centered at $\lambda=1$:
	\begin{equation}
	\pi(\lambda) \propto \exp \left( - \frac{\left(\lambda-1\right)^2}{2\epsilon^2} \right).
	\end{equation}
The standard deviation $\epsilon$ encapsulates the amplitude calibration uncertainty.
Within the 20-1726 Hz frequency band, LIGO-Hanford and LIGO-Livingston have maximum estimated amplitude uncertainties of $\blue{4.8}\%$ and $\blue{5.4}\%$, respectively \cite{Cahillane2017}.
These uncertainty estimates have been improved relative to the the uncertainties previously adopted in Refs. \cite{IsotropicO1,IsotropicO1Supplement,DirectionalO1}.
For our analysis, we take $\epsilon = \blue{0.072}$, the quadrature sum of the Hanford and Livingston uncertainties \cite{IsotropicO1}.
All results are given after marginalization over $\lambda$.

In the above prescription we have made two simplifying assumptions.
First, we have neglected phase calibration uncertainty, which is expected to be a sub-dominant source of uncertainty in the stochastic analysis \cite{Cahillane2017,Whelan2014}.
Secondly, although calibration uncertainties are frequency-dependent, for simplicity we adopt uniform amplitude uncertainties across all frequencies.
Our quoted amplitude uncertainties are conservative, encompassing the largest calibration uncertainties in the stochastic sensitivity band \cite{IsotropicO1,Cahillane2017}.

\section{Detailed Parameter Estimation Results}

\begin{figure*}
\centering
\includegraphics[width=\textwidth]{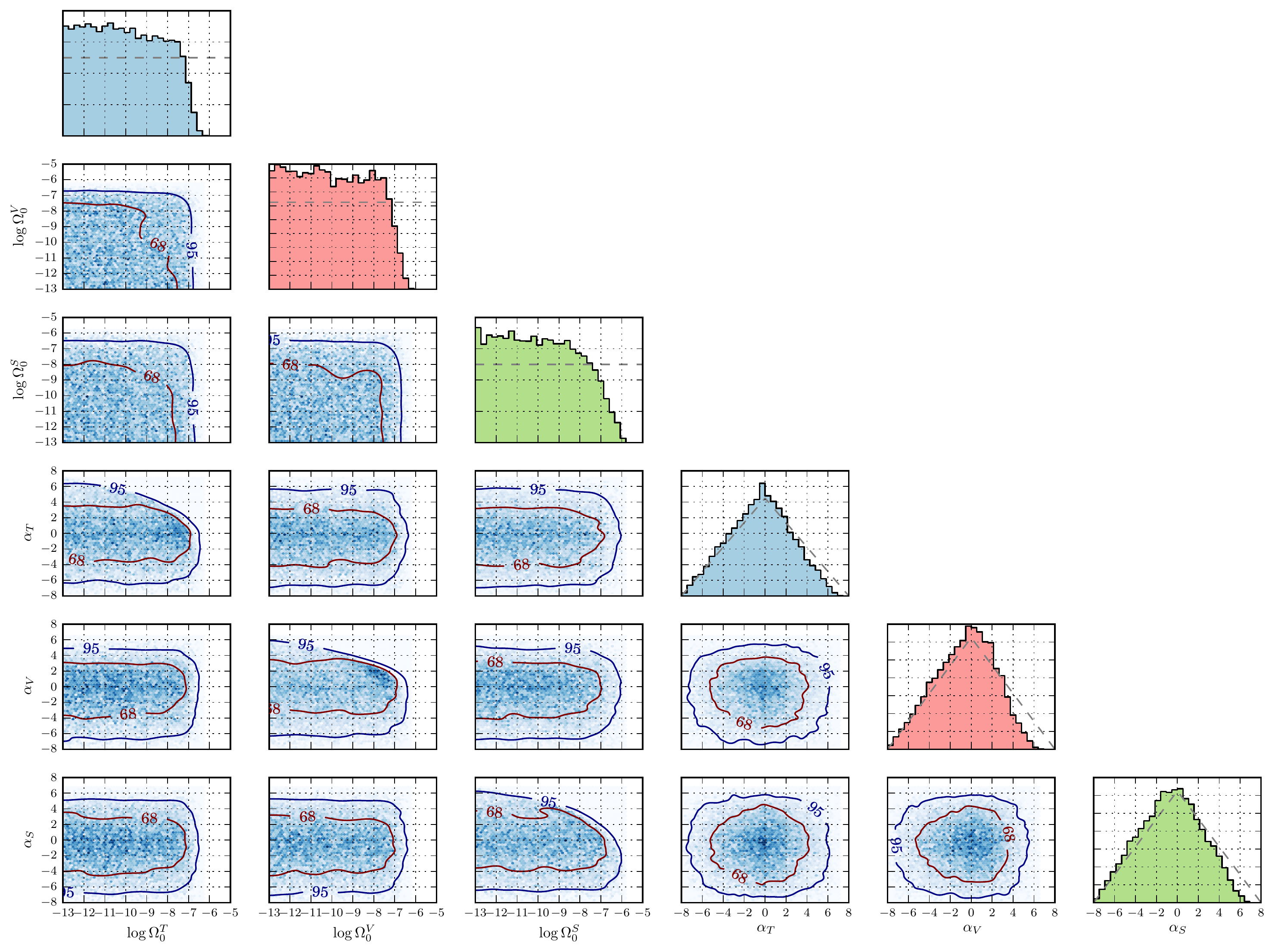}
\caption{
Posterior probability distributions for the power-law amplitudes and slopes of tensor, vector, and scalar contributions to the stochastic background, assuming the TVS hypothesis and log-uniform amplitude priors.
The subplots along the diagonal show marginalized posteriors for each parameter; dashed curves show the corresponding prior.
The marginalized amplitude posteriors yield the 95\% credible upper limits given in Table \ref{peTable}.
The remaining subplots, meanwhile, show the two-dimensional posteriors between each pair of parameters, as well as contours containing the central 68\% and 95\% posterior probability.
}
\label{TVS_logUniform}
\end{figure*}

In our Letter, we presented marginalized posteriors for the tensor, vector, and scalar background amplitudes under the ``TVS'' hypothesis.
In Fig. \ref{TVS_logUniform} we show the full six-dimensional parameter estimation results obtained when choosing log-uniform amplitude priors.
Diagonal subplots show marginalized posteriors on the amplitudes and slopes of each polarization, while the interior subplots show joint posteriors between each pair of parameters.
The spectral index posteriors (not shown in the main Letter) are largely consistent with our choice of prior, but indicate a slight bias against large positive spectral indices.
This reflects the fact that Advanced LIGO is most sensitive to backgrounds of large, positive slopes \cite{Callister2017}.
The non-detection of a stochastic background therefore constrains larger amplitudes to have small and/or negative spectral indices; see, for instance, the joint $\log\Omega^T_0\text{-- }\alpha_T$ posterior in Fig. \ref{TVS_logUniform}.

\begin{figure*}
\centering
\includegraphics[width=\textwidth]{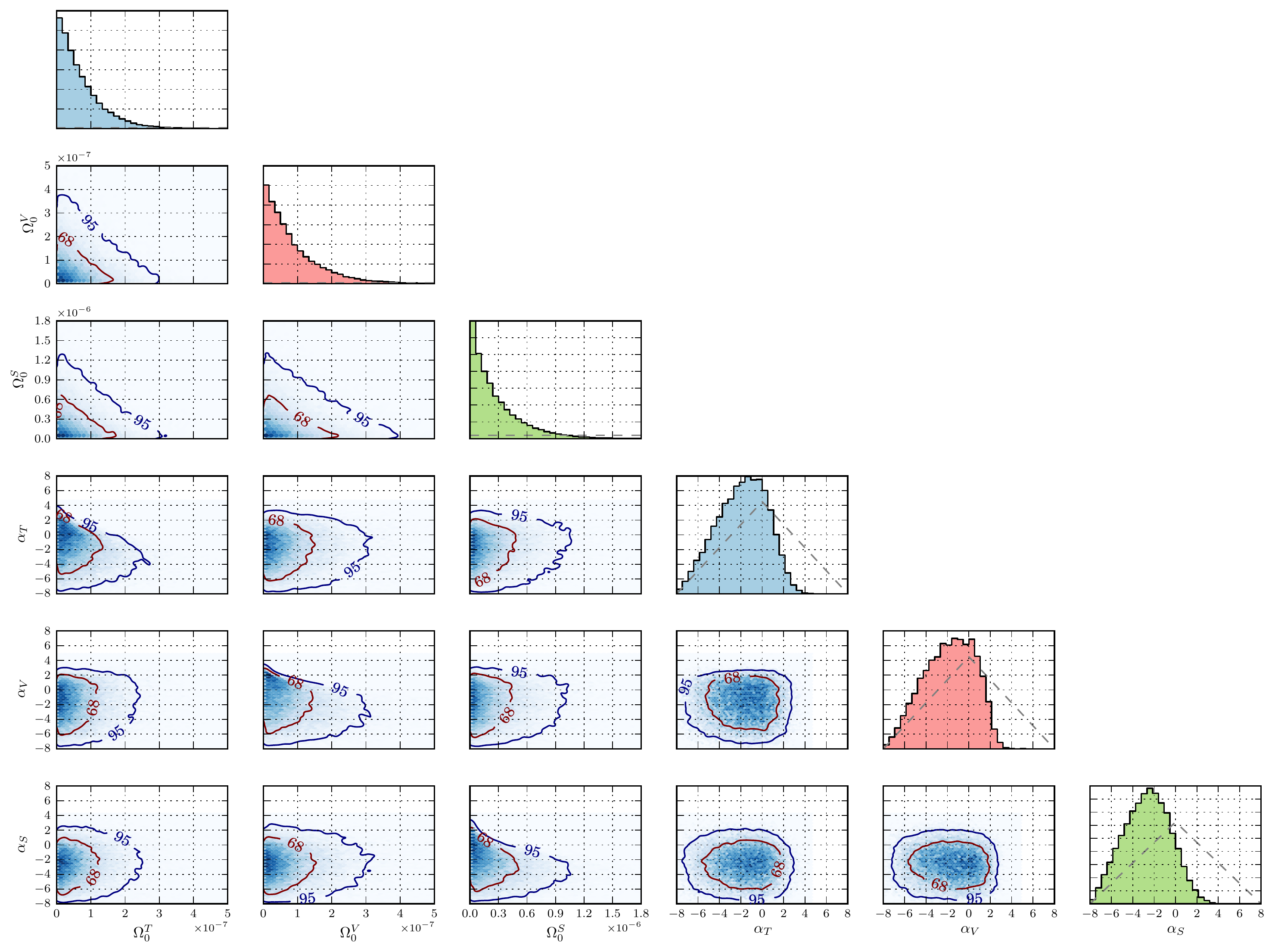}
\caption{
As in Fig. \ref{TVS_uniform}, but assuming a uniform prior on background amplitudes.
The marginalized amplitude posteriors yield the 95\% credible upper limits given in Table \ref{peTable}.
}
\label{TVS_uniform}
\end{figure*}

Figure \ref{TVS_uniform}, meanwhile, shows full parameter estimation results when alternatively assuming uniform amplitude priors.
Here, the posterior preference towards small or negative spectral indices is far more pronounced.
The joint 2D posteriors (e.g. $\Omega^T_0\text{-- }\alpha_T$) again illustrate that large, positive slopes are preferentially ruled out in case of large background amplitudes.

As stated in the Letter, upper limits obtained under one hypothesis are not, in general, equal to those obtained under some different hypothesis.
While we presented upper limits only for the TVS hypothesis, results from other hypotheses may be desired as well (the TS results, for instance, are best suited for comparison to predictions from scalar-tensor theories).
In Tables \ref{fullPE_logUniform} and \ref{fullPE_uniform} we have therefore listed the 95\% credible upper limits corresponding to each signal sub-hypothesis, for both log-uniform and uniform amplitude priors.
We have also listed 95\% credible bounds on spectral indices for each choice of amplitude prior.

\begin{table*}[!h]
\caption{
Parameter estimation results for each signal sub-hypothesis, obtained with log-uniform priors on the amplitude of each polarization component.
Columns 2-4 give 95\% credible upper limits on $\log\Omega^A_0$, and columns 5-7 show the corresponding limits on $\Omega^A_0$ for convenience.
Columns 8-10, meanwhile, show the central 95\% credible bounds on spectral indices $\alpha_A$.
The parameter estimation results for the TVS sub-hypothesis (final row) correspond to those given in Table \ref{peTable} of the main text.
}
\label{extraPEtable_logUniform}
\setlength{\tabcolsep}{3pt}
\renewcommand{\arraystretch}{1.4}
\begin{ruledtabular}
\begin{tabular}{l | r r r  | r r r | r r r}
Hypothesis & $\log\Omega^{T,95\%}_0$ & $\log\Omega^{V,95\%}_0$ & $\log\Omega^{S,95\%}_0$ 
	& $\Omega^{T,95\%}_0$ & $\Omega^{V,95\%}_0$ & $\Omega^{S,95\%}_0$ 
	& $\alpha_T$ & $\alpha_V$ & $\alpha_S$ \\
\hline
T & $\logOmgTlogUniformT$ 
	& -
	& -
	& $\num{\OmgTlogUniformT}$
	& -
	& -
	& $\alphaTlogUniformTmed^{+\alphaTlogUniformTupper}_{-\alphaTlogUniformTlower}$ 
	& - 
	& - \\
V & - 
	& $\logOmgVlogUniformV$ 
	& - 
	& - 
	& $\num{\OmgVlogUniformV}$
	& -
	& -
	& $\alphaVlogUniformVmed^{+\alphaVlogUniformVupper}_{-\alphaVlogUniformVlower}$ 
	& - \\ 
S & - 
	& - 
	& $\logOmgSlogUniformS$ 
	& - 
	& - 
	& $\num{\OmgSlogUniformS}$
	& -
	& -
	& $\alphaSlogUniformSmed^{+\alphaSlogUniformSupper}_{-\alphaSlogUniformSlower}$ \\ 
TV & $\logOmgTlogUniformTV$ 
	& $\logOmgVlogUniformTV$ 
	& - 
	& $\num{\OmgTlogUniformTV}$
	& $\num{\OmgVlogUniformTV}$
	& -
	& $\alphaTlogUniformTVmed^{+\alphaTlogUniformTVupper}_{-\alphaTlogUniformTVlower}$ 
	& $\alphaVlogUniformTVmed^{+\alphaVlogUniformTVupper}_{-\alphaVlogUniformTVlower}$ 
	& - \\
TS & $\logOmgTlogUniformTS$ 
	& -
	& $\logOmgSlogUniformTS$ 
	& $\num{\OmgTlogUniformTS}$
	& -
	& $\num{\OmgSlogUniformTS}$
	& $\alphaTlogUniformTSmed^{+\alphaTlogUniformTSupper}_{-\alphaTlogUniformTSlower}$ 
	& -
	& $\alphaSlogUniformTSmed^{+\alphaSlogUniformTSupper}_{-\alphaSlogUniformTSlower}$ \\
VS & - 
	& $\logOmgVlogUniformVS$ 
	& $\logOmgSlogUniformVS$ 
	& -
	& $\num{\OmgVlogUniformVS}$
	& $\num{\OmgSlogUniformVS}$
	& -
	& $\alphaVlogUniformVSmed^{+\alphaVlogUniformVSupper}_{-\alphaVlogUniformVSlower}$
	& $\alphaSlogUniformVSmed^{+\alphaSlogUniformVSupper}_{-\alphaSlogUniformVSlower}$ \\
TVS & $\logOmgTlogUniformTVS$ 
	& $\logOmgVlogUniformTVS$ 
	& $\logOmgSlogUniformTVS$ 
	& $\num{\OmgTlogUniformTVS}$
	& $\num{\OmgVlogUniformTVS}$
	& $\num{\OmgSlogUniformTVS}$
	& $\alphaTlogUniformTVSmed^{+\alphaTlogUniformTVSupper}_{-\alphaTlogUniformTVSlower}$
	& $\alphaVlogUniformTVSmed^{+\alphaVlogUniformTVSupper}_{-\alphaVlogUniformTVSlower}$
	& $\alphaSlogUniformTVSmed^{+\alphaSlogUniformTVSupper}_{-\alphaSlogUniformTVSlower}$ \\
\end{tabular}
\end{ruledtabular}
\label{fullPE_logUniform}
\end{table*}

\begin{table*}[ht!]
\caption{
Parameter estimation results for each signal sub-hypothesis, obtained with uniform priors on the background amplitudes.
Columns are defined as in Table \ref{extraPEtable_logUniform} above.
The parameter estimation results for the TVS sub-hypothesis (final row) correspond to those given in Table \ref{peTable} of the main text.
}
\label{extraPEtable_uniform}
\setlength{\tabcolsep}{3pt}
\renewcommand{\arraystretch}{1.4}
\begin{ruledtabular}
\begin{tabular}{l | r r r  | r r r | r r r}
Hypothesis & $\log\Omega^{T,95\%}_0$ & $\log\Omega^{V,95\%}_0$ & $\log\Omega^{S,95\%}_0$ 
	& $\Omega^{T,95\%}_0$ & $\Omega^{V,95\%}_0$ & $\Omega^{S,95\%}_0$ 
	& $\alpha_T$ & $\alpha_V$ & $\alpha_S$ \\
\hline
T & $\logOmgTuniformT$ 
	& -
	& -
	& $\num{\OmgTuniformT}$
	& -
	& -
	& $\alphaTuniformTmed^{+\alphaTuniformTupper}_{-\alphaTuniformTlower}$ 
	& - 
	& - \\
V & - 
	& $\logOmgVuniformV$ 
	& - 
	& - 
	& $\num{\OmgVuniformV}$
	& -
	& -
	& $\alphaVuniformVmed^{+\alphaVuniformVupper}_{-\alphaVuniformVlower}$ 
	& - \\ 
S & - 
	& - 
	& $\logOmgSuniformS$ 
	& - 
	& - 
	& $\num{\OmgSuniformS}$
	& -
	& -
	& $\alphaSuniformSmed^{+\alphaSuniformSupper}_{-\alphaSuniformSlower}$ \\ 
TV & $\logOmgTuniformTV$ 
	& $\logOmgVuniformTV$ 
	& - 
	& $\num{\OmgTuniformTV}$
	& $\num{\OmgVuniformTV}$
	& -
	& $\alphaTuniformTVmed^{+\alphaTuniformTVupper}_{-\alphaTuniformTVlower}$ 
	& $\alphaVuniformTVmed^{+\alphaVuniformTVupper}_{-\alphaVuniformTVlower}$ 
	& - \\
TS & $\logOmgTuniformTS$ 
	& -
	& $\logOmgSuniformTS$ 
	& $\num{\OmgTuniformTS}$
	& -
	& $\num{\OmgSuniformTS}$
	& $\alphaTuniformTSmed^{+\alphaTuniformTSupper}_{-\alphaTuniformTSlower}$ 
	& -
	& $\alphaSuniformTSmed^{+\alphaSuniformTSupper}_{-\alphaSuniformTSlower}$ \\
VS & - 
	& $\logOmgVuniformVS$ 
	& $\logOmgSuniformVS$ 
	& -
	& $\num{\OmgVuniformVS}$
	& $\num{\OmgSuniformVS}$
	& -
	& $\alphaVuniformVSmed^{+\alphaVuniformVSupper}_{-\alphaVuniformVSlower}$
	& $\alphaSuniformVSmed^{+\alphaSuniformVSupper}_{-\alphaSuniformVSlower}$ \\
TVS & $\logOmgTuniformTVS$
	& $\logOmgVuniformTVS$ 
	& $\logOmgSuniformTVS$ 
	& $\num{\OmgTuniformTVS}$
	& $\num{\OmgVuniformTVS}$
	& $\num{\OmgSuniformTVS}$
	& $\alphaTuniformTVSmed^{+\alphaTuniformTVSupper}_{-\alphaTuniformTVSlower}$
	& $\alphaVuniformTVSmed^{+\alphaVuniformTVSupper}_{-\alphaVuniformTVSlower}$
	& $\alphaSuniformTVSmed^{+\alphaSuniformTVSupper}_{-\alphaSuniformTVSlower}$ \\
\end{tabular}
\end{ruledtabular}
\label{fullPE_uniform}
\end{table*}